\documentclass[12pt]{article}
\usepackage{amsmath}
\usepackage{amsfonts}
\usepackage{amssymb}
\usepackage{amsthm}
\usepackage{cite}

\newcommand{\Ppr}{P_{(p,\rad)} }
\newcommand{\rad}{r}

\newcommand{\e}{\hat{e}}

\newcommand{\lam}{t}

\newcommand{\omex}{\om_0^+}
\newcommand{\h}{\half}

\newcommand{\SiEh}{\Pi_E}

\newcommand{\lattice}{\mathbb{J}}
\newcommand{\khat}{\hat{k}}

\newcommand{\vxxb}{\underline{x}}
\newcommand{\vxz}{\underline{z}}

\newcommand{\N}{\mathcal{N}}
\newcommand{\Sa}{T}
\newcommand{\s}{t}
\newcommand{\T}{S}
\newcommand{\bsharp}{\mbox{\boldmath $^\sharp$}}
\newcommand{\bnatural}{\mbox{\boldmath $^\natural$}}
\newcommand{\p}{*}
\newcommand{\cbe}{(2\sqrt{\pi}\be)^{-s} }
\newcommand{\reals}{\real^{s+1}}

\newcommand{\Cs}{C_{\bsharp}}
\newcommand{\Ns}{N_{\bsharp}}
\newcommand{\x}{x_1,\ldots, x_N}
\newcommand{\oA}{\mathring{A}}

\newcommand{\M}{M_E}

\newcommand{\damp}{e^{-\half(\be|\vep|)^2}}
\newcommand{\dam}{e^{-(\be|\vep|)^2} }
\newcommand{\pin}{\fr{1}{p}}

\newcommand{\Thet}{{\Theta}_{E,\vxb}}
\newcommand{\Theh}{\Theta_{E,\vxb}}

\newcommand{\lin}{\mathcal{L}}

\newcommand{\vy}{\vec{y}}

\newcommand{\PiE}{\Pi_E}
\newcommand{\al}{\alpha}
\newcommand{\be}{\beta}
\newcommand{\La}{\Lambda}
\newcommand{\la}{\lambda}
\newcommand{\vp}{\varphi}
\newcommand{\Ga}{\Gamma}

\newcommand{\om}{\omega}
\newcommand{\eps}{\varepsilon}
\newcommand{\De}{\Delta}
\newcommand{\de}{\delta}
\newcommand{\The}{\Theta}
\newcommand{\SEp}{\Sa_{E,+}}
\newcommand{\SEm}{\Sa_{E,-}}
\newcommand{\SEpm}{\Sa_{E,\pm}}

\newcommand{\Sbpm}{\Sa_{\be,\pm}}
\newcommand{\Sbp}{\Sa_{\be,+}}
\newcommand{\Sbm}{\Sa_{\be,-}}
\newcommand{\LL}{\mathcal{L}}
\newcommand{\Lp}{\mathcal{L}^{+}}
\newcommand{\Lm}{\mathcal{L}^{-}}
\newcommand{\Lpm}{\mathcal{L}^{\pm}}
\newcommand{\LJ}{\mathcal{L}}
\newcommand{\integer}{\mathbb{Z}}

\newcommand{\nat}{\mathbb{N}}

\newcommand{\real}{\mathbb{R}}
\newcommand{\complex}{\mathbb{C}}
\newcommand{\mup}{\mu^+}
\newcommand{\mum}{\mu^-}

\newcommand{\mub}{\overline{\mu}}
\newcommand{\nub}{\overline{\nu}}
\newcommand{\nuh}{\nub_a}
\newcommand{\nuc}{\nub_b}
\newcommand{\Bnh}{B_{\nuh}}
\newcommand{\Bnc}{B_{\nuc}}
\newcommand{\Bm}{B_{\mub}}
\newcommand{\Bn}{B_{\nub}}
\newcommand{\Tmn}{\T_{\mub,\nub}}
\newcommand{\trace}{\mathcal{T}}
\newcommand{\traceE}{\trace_E}

\newcommand{\traceEB}{\trace_{E,1}}
\newcommand{\traceEBP}{\trace_{E,1}^+}
\newcommand{\trac}{\mathring{\trace}_{E}}
\newcommand{\track}{\mathring{\trace}_{E_k,1}}
\newcommand{\trackk}{\mathring{\trace}_{E_k}}
\newcommand{\tracB}{\mathring{\trace}_{E,1}}

\newcommand{\traceprc}{\mathring{\trace}_{(p,\rad)} }
\newcommand{\traceprcz}{\mathring{\trace}_{(0,\sqrt{2}E)} }

\newcommand{\vxb}{\x}
\newcommand{\vep}{\vec{p}}
\newcommand{\veP}{\vec{P}}

\newcommand{\vx}{\vec{x}}
\newcommand{\Nnat}{N_{\bnatural}}

\newcommand{\bdot}{\: \cdot \:}

\newcommand{\compsup}{\complex^N_{\sup}}

\newcommand{\funr}{\fun_{\textrm{Re}}}
\newcommand{\funi}{\fun_{\textrm{Im}}}
\newcommand{\fun}{\vp}

\newcommand{\su}{\substack}

\newcommand{\hil}{\mathcal{H}}
\newcommand{\mfa}{\mathfrak{A}}
\newcommand{\mco}{\mathcal{O}}

\newcommand{\cone}{\overline{V}_+}

\newcommand{\supp}{\textrm{supp}}
\newcommand{\fr}[2]{\frac{#1}{#2}}
\newcommand{\ot}{\otimes}

\newcommand{\non}{\nonumber}

\newcommand{\vac}{\Omega}
\newcommand{\fp}{f^+}
\newcommand{\fm}{f^-}
\newcommand{\half}{\fr{1}{2}}
\newcommand{\lan}{\langle}
\newcommand{\ran}{\rangle}

\def\proof{\noindent\emph{Proof. }}
\def\qed{$\Box$\medskip}

\newtheorem{theoreme}{Theorem } [section]
\newtheorem{proposition}[theoreme]{Proposition}
\newtheorem{lemma}[theoreme]{Lemma}
\newtheorem{definition}[theoreme]{Definition}
\newtheorem{corollary}[theoreme]{Corollary}
\newtheorem{remark}[theoreme]{Remark}
\newtheorem{example}[theoreme]{Example}
\newtheorem{criterion}[theoreme]{Criterion}

\newcommand{\beq}{\begin{equation}}
\newcommand{\eeq}{\end{equation}}
\newcommand{\beqa}{\begin{eqnarray}}
\newcommand{\eeqa}{\end{eqnarray}}
\newcommand{\ben}{\begin{arabicenumerate}}
\newcommand{\een}{\end{arabicenumerate}}
\newcommand{\bex}{\begin{example}}
\newcommand{\eex}{\end{example}}
\newcommand{\ber}{\begin{remark}}
\newcommand{\eer}{\end{remark}}
\newcommand{\bec}{\begin{corollary}}
\newcommand{\eec}{\end{corollary}}
\newcommand{\bep}{\begin{proposition}}
\newcommand{\eep}{\end{proposition}}
\newcommand{\becr}{\begin{criterion}}
\newcommand{\eecr}{\end{criterion}}

\def\bel{\begin{lemma}}
\def\eel{\end{lemma}}
\def\bet{\begin{theoreme}}
\def\eet{\end{theoreme}}
\def\bed{\begin{definition}}
\def\eed{\end{definition}}

\begin{document}

\title{A sharpened nuclearity condition and the uniqueness of the vacuum in QFT}
\author{Wojciech Dybalski\\ [5mm] 
  Institut f\"ur Theoretische Physik, Universit\"at G\"ottingen, \\[2mm]
Friedrich-Hund-Platz 1, D-37077 G\"ottingen - Germany \\ [2mm]
 e-mail: dybalski@theorie.physik.uni-goettingen.de}
\date{}
\maketitle
\begin{abstract}
It is shown that only one vacuum state can be prepared with a finite amount of energy and
it appears, in particular, as a limit of physical states under large timelike translations in any theory
which satisfies a phase space condition proposed in this work. This new criterion, related to the 
concept of additivity of energy over isolated subsystems, is verified in massive free field theory. The analysis entails
very detailed results about the momentum transfer of local operators in this model.

\end{abstract}

\section{Introduction}
\setcounter{equation}{0}
Since the seminal work of Haag and Swieca \cite{HS} restrictions on the phase space 
structure of a theory formulated in terms of compactness and nuclearity conditions have proved very useful in the structural analysis of quantum field theories \cite{BW,BJ,Scaling,Bos1,Bos2} and in the
construction of interacting models \cite{Gan, Gan1}. However, the initial goal
of Haag and Swieca, namely to characterize theories which have a reasonable particle interpretation, has not 
been accomplished to date. While substantial progress was made in our understanding of the timelike 
asymptotic behavior of physical states \cite{AH,Enss,Buch3,BPS,Porr1,Porr2,BS}, several important
convergence and existence questions remained unanswered. As a matter of fact, it turned out that the original compactness condition introduced in \cite{HS} is not sufficient to settle these issues.

Therefore, in the present article we propose a sharpened phase space condition, stated below, which seems 
to be more appropriate. We show that it is related to additivity of energy over isolated subregions and 
implies that there is only one vacuum state within the energy-connected component  of the state space,
as one expects in physical spacetime \cite{BWa}. We stress that there may exist other vacua in a theory complying with our condition, but, loosely speaking, they are separated by an infinite energy barrier and thus not accessible to experiments.
The convergence of physical states to the vacuum state under large timelike translations is a corollary of
this discussion. A substantial part of this work is devoted to the proof that the new condition holds in massive scalar free field theory. As a matter of fact, it holds also in the massless case which will be treated elsewhere.
These last results demonstrate that the new criterion is consistent with the basic postulates of local relativistic quantum field theory \cite{Haag} which we now briefly recall.

The theory is based on a local net $\mco\to\mfa(\mco)$ of von Neumann algebras, which are attached to open, 
bounded regions of spacetime $\mco\subset\real^{s+1}$ and act on a Hilbert space $\hil$. The global algebra of 
this net, denoted by $\mfa$, is irreducibly represented on this space. Moreover, $\hil$ carries a strongly continuous unitary representation of the Poincar\'e group $\real^{s+1}\rtimes L_+^{\uparrow}\ni (x,\La)\to U(x,\La)$ which acts geometrically on the net
\beq
\al_{(x,\La)}\mfa(\mco)=U(x,\La)\mfa(\mco)U(x,\La)^{-1}=\mfa(\La\mco+x).
\eeq
We adopt the usual notation for translated operators $\al_x A=A(x)$ and functionals $\al_x^\p\fun(A)=\fun(A(x))$,
where $A\in \mfa$,  $\fun\in \mfa^*$, and
demand that the joint spectrum of the generators of translations $H, P_1,\ldots ,P_s$ 
is contained in the closed forward lightcone~$\cone$.
We denote by $P_E$ the spectral projection of $H$ (the Hamiltonian) on the subspace spanned by vectors of energy lower than $E$.
Finally, we identify the predual of $B(\hil)$ with the space $\trace$ of trace-class operators on $\hil$
and  denote by $\traceE=P_E\trace P_E$ the space of normal functionals of energy bounded by $E$. We assume that
there exists a vacuum state $\om_0\in\traceE$ and introduce the subspace $\trac=\{\fun-\fun(I)\om_0 \ | \ \fun\in\traceE\}$ of functionals with the asymptotically dominant vacuum contribution subtracted.

The main object of our investigations is the family of maps $\PiE: \trac\to \mfa(\mco)^*$ given by
\beq
\PiE(\fun)=\fun|_{\mfa(\mco)},\quad \fun\in\trac.
\eeq
Fredenhagen and Hertel argued in some unpublished work that in physically meaningful theories these maps should be subject to the following restriction:
\begin{enumerate}
\item[] \bf Condition \rm $\Cs$. The maps $\PiE$ are compact for any $E\geq 0$ and  
double cone $\mco\subset\real^{s+1}$.
\end{enumerate}
This condition is expected to hold in theories exhibiting mild infrared behavior \cite{BP}.
In order to restrict the number of local degrees of freedom also in the ultraviolet part of 
the energy scale, Buchholz and Porrmann proposed a stronger condition  which makes use of the concept of nuclearity\footnote{We recall that a map $\Pi: X\to Y$ is $p$-nuclear if there 
exists a decomposition $\Pi=\sum_n\Pi_n$ into rank-one maps s.t. $\nu^p:=\sum_n\|\Pi_n\|^p<\infty$.
The $p$-norm $\|\Pi\|_p$ of this map is the smallest such $\nu$ and it is equal to zero for $p>1$ \cite{FOP}. 
Note that for any norm on $\lin(X,Y)$ one can introduce the corresponding class of $p$-nuclear  maps.
Similarly, we say that a map is compact w.r.t. a given norm on $\lin(X,Y)$  if it can be approximated 
by finite rank mappings in this norm. } \cite{BP}:
\begin{enumerate}
\item[] \bf Condition \rm $\Ns$. The maps $\PiE$ are $p$-nuclear for any $0<p\leq 1$, $E\geq 0$ and  
double cone $\mco\subset\real^{s+1}$.
\end{enumerate} 
This condition is still somewhat conservative since it does not take into account
the fact that for any $\fun\in\trac$ the restricted functionals $\al_x^\p\fun|_{\mfa(\mco)}$ should be arbitrarily close to zero apart from translations varying in some compact subset of $\real^{s+1}$, depending on $\fun$. It seems therefore desirable to introduce a family of norms on $\lin(\trac,X)$, where $X$ is some Banach space, given for any $N\in\nat$ and $\x\in\real^{s+1}$ by
\beq
\|\Pi\|_{\x}=\sup_{\fun\in\tracB}\bigg(\sum_{k=1}^N\|\Pi(\al_{x_k}^\p\fun)\|^2\bigg)^{\half},\quad \Pi\in \lin(\trac,X), \label{Nnorm}
\eeq
and the corresponding family of $p$-norms $\|\Pi\|_{p,\x}$, (see footnote 1).
It is easily seen that if $\PiE$ satisfies Condition $\Cs$, respectively $\Ns$, then $\PiE$ is also compact,
respectively $p$-nuclear, with respect to the above norms, and vice versa. Important additional information 
is contained in the dependence of the nuclear $p$-norms on $N$.
In Sect. 2 we argue that the natural assumption is:
\begin{enumerate}
\item[] \bf Condition \rm $\Nnat$. The maps $\PiE$ are $p$-nuclear w.r.t. the norms $\|\cdot~\|_{\x}$
for any $N\in\nat$, $\x\in\real^{s+1}$,  $0<p\leq 1$, $E\geq 0$
and double cone $\mco\subset\reals$. Moreover, there holds for their nuclear $p$-norms
\beq
\limsup\|\PiE\|_{p,\x}\leq c_{p,E}, \label{natbound}
\eeq
where $c_{p,E}$ is independent of $N$ and the limit is taken for configurations $\x$, where all $x_i-x_j$, $i\neq j$, tend to spacelike infinity.
\end{enumerate}
Restricting attention to the case $N=1$, it is easily seen that Condition $\Nnat$ implies Condition $\Ns$,
but not vice versa. 

Our paper is organized as follows: In Sect. 2 we show that Condition $\Nnat$ implies a certain form of additivity of energy over isolated subsystems and  guarantees the physically meaningful vacuum structure of a theory. More technical part of this discussion
is postponed to Appendix A. In Sect. 3 we recall some basic facts about  massive scalar free field theory and its phase space structure. In Appendix B we provide a simple proof of the known fact that Condition $\Ns$ holds in this model. 
Sect. 4 contains our main technical result, namely the proof that  Condition $\Nnat$ holds in this theory as well. The argument demonstrates, in this simple example, the interplay between locality and positivity of energy which allows to strengthen Condition $\Ns$. The paper concludes with a brief outlook where we apply our techniques to the harmonic analysis of translation automorphisms.

\section{Physical Consequences of Condition $\Nnat$}
\setcounter{equation}{0}
In this section we show that theories satisfying Condition $\Nnat$ exhibit two physically desirable properties:
a variant of additivity of energy over isolated subregions and the feature that only one vacuum state can be prepared
given a finite amount of energy. Combining this latter property with covariance of a theory under Lorentz transformations
we will conclude that physical states converge to the vacuum state under large timelike translations.

The concept of additivity of energy over isolated subsystems does not have an unambiguous meaning in the general framework of local relativistic quantum field theory and we rely here on the following formulation:
We introduce the family of maps $\Thet: \trac\to \mfa(\mco)^*\ot\compsup$,  given by
\beq
\Thet(\fun)=\big(\PiE(\al_{x_1}^*\fun),\ldots,\PiE(\al_{x_N}^*\fun) \big), \label{theta}
\eeq
where $\compsup$ denotes the space $\complex^N$ equipped with the norm $\|\vxz\|=\sup_{k\in\{1,\ldots,N\}}|z_k|$.
We claim that a mild (polynomial) growth of the $\eps$-contents\footnote{The $\eps$-content of a map $\Pi: X\to Y$ is the maximal natural number $\N(\eps)$ for which there exist elements $\fun_1,\ldots,\fun_{\N(\eps)}\in X_1$ s.t. $\|\Pi(\fun_i)-\Pi(\fun_j)\|>\eps$ for $i\neq j$. Clearly, $\N(\eps)$ is finite for any $\eps>0$ if the map $\Pi$ is compact.}  $\N(\eps)_{E,\vxb}$ of these maps with $N$, (when $x_i-x_j$, $i\neq j$, tend to spacelike infinity),
is a signature of additivity of energy over isolated subregions. In order to justify this formulation
we provide a heuristic argument: Given a functional $\fun\in\tracB$, we denote by $E_k$ the 'local energy content' of the restricted functional $\fun|_{\mfa(\mco+x_k)}$. Additivity of energy should then imply that $E_1+\cdots+E_N\leq E$ for large spacelike distances between the regions $\mco+x_1,\ldots,\mco+x_N$. This suggests that to calculate
$\N(\eps)_{E,\vxb}$ one should count all the families of  functionals
$(\fun_1,\ldots,\fun_N)$, $\fun_k\in\track$, $E_1+\cdots+E_N\leq E$, which can be distinguished, up to
accuracy $\eps$, by measurements in $\mco+x_1,\ldots,\mco+x_N$. Relying on this heuristic reasoning we write
\beqa
\N(\eps)_{E,\vxb}=\#\{\, (n_1\ldots n_N)\in\nat^{*\times N}\, |\,
n_1\leq \N(\eps)_{E_1},\ldots, n_N \leq\N(\eps)_{E_N}, \non\\  
\textrm{ for some } E_1,\ldots,E_N\geq 0\textrm{ s.t. }  E_1+\cdots+E_N\leq E \,\}, \label{heuristic}
\eeqa
where we made use of the fact that  the number of functionals from $\track$ which can be discriminated, up to $\eps$, by observables localized in the region $\mco+x_k$ is equal to the $\eps$-content $\N(\eps)_{E_k}$ of the map $\Pi_{E_k}:\trackk\to\mfa(\mco+x_k)$ given by $\Pi_{E_k}(\fun)=\fun|_{\mfa(\mco+x_k)}$. 
Anticipating that $\N(\eps)_{E_k}$ tends to one for small $E_k$ we may assume that 
\beq
\N(\eps)_{E_k}\leq 1+c(\eps, E)E_k \label{ff}
\eeq
for $E_k\leq E$. (This is valid e.g. in free field theory due to  Sect. 7.2 of \cite{Bos3} and  Proposition 2.5 iii of \cite{BA}).  From the heuristic formula (\ref{heuristic}) and the bound (\ref{ff}) we obtain the estimate which grows only polynomially with $N$
\beqa
\N(\eps)_{E,\vxb}\leq\#\{\, (n_1\ldots n_N)\in\nat^{*\times N}\, |\, n_1+\cdots+n_N\leq N+c(\eps,E)E\, \}\non\\
\phantom{444444444444444444444444444444444444444444444444444}\leq (N+1)^{c(\eps,E)E},
\eeqa
where the last inequality can be verified by induction in $N$. Omitting the key condition $E_1+\cdots+E_N\leq E$ in (\ref{heuristic}) and setting  $E_k=E$ instead, one would arrive at an exponential growth of $\N(\eps)_{E,\vxb}$ as a function of $N$. Thus the moderate (polynomial) increase of this quantity with regard to $N$ is in fact a clear-cut signature of additivity of energy over isolated subsystems. It is therefore of interest that this feature prevails in all theories complying with Condition~$\Nnat$ as shown in the subsequent theorem whose
proof is given in Appendix~A.
\bet\label{Nregions} Suppose that Condition $\Nnat$ holds. Then the $\eps$-content 
$\N(\eps)_{E,\vxb}$ of the map $\Thet$ satisfies
\beq
\limsup\N(\eps)_{E,\vxb}\leq (4eN)^{\fr{c(E)}{\eps^2}}, \label{addenergy}
\eeq
where the constant $c(E)$ is independent of $N$ and the limit is taken for configurations 
$\x$, where all $x_i-x_j$, $i\neq j$, tend to spacelike infinity.
\eet
Now let us turn our attention to the vacuum structure of the theories under study. In physical spacetime one expects that there is a unique vacuum state which can be prepared with a finite amount of energy. This fact is related to additivity of energy and
can be derived from Condition~$\Nnat$.
\bet\label{disconnected} Suppose that a state $\om\in\mfa^*$ belongs to the weak* closure of $\traceEB$ for some $E\geq 0$ and is invariant under translations along some spacelike ray. Then the following assertions hold:
\begin{enumerate}
\item[(a)] If Condition $\Cs$ is satisfied, $\om$ is a vacuum state.
\item[(b)] If Condition $\Nnat$ is satisfied, $\om$ coincides with the vacuum state $\om_0$.
\end{enumerate}
\eet
\proof 
(a) We pick any $A\in\mfa(\mco)$, a test function $f\in S(\real^{s+1})$ s.t. $\supp\tilde{f}\cap\cone=\emptyset$
and define the energy decreasing operator $A(f)=\int A(x) f(x) d^{s+1}x$.
Next, we parametrize the ray from the statement of the theorem as $\{ \ \la \e \ | \ \la\in\real \ \}$, 
where $\e\in\real^{s+1}$ is some spacelike unit vector, choose  a compact subset $K\subset\real$ and estimate
\beqa
\om(A(f)^*A(f))|K|&=&\int_{K}d\la \ \om\big( (A(f)^*A(f))(\la \e) \big)\non\\
&=&\lim_{n\to\infty}\fun_n\bigg(\int_{K}d\la \ (A(f)^*A(f))(\la \e) \bigg)\non\\
&\leq& \|P_{E}\int_{K} d\la \ (A(f)^*A(f))(\la \e) \ P_{E}\|. \label{harmonic1}
\eeqa
In the first step we exploited invariance of the state $\om$ under translations along the spacelike ray. In
 the second step we made use of local normality of this state, which follows from Condition $\Cs$, in order to exchange its action with integration. Approximating $\om$ by a sequence of functionals $\fun_n\in\traceEB$,
we arrived at the last expression. (Local normality of $\om$ and existence of an approximating sequence
can be shown as in \cite{GJ} p. 49).
Now we can apply a slight modification of Lemma 2.2 from \cite{Buch3}, (see also Lemma \ref{harmonic} below), to conclude that the last expression on the r.h.s. of (\ref{harmonic1}) is bounded uniformly in $K$. 
As $|K|$ can be made arbitrarily large, it follows that
\beq
\om(A(f)^*A(f))=0 \label{zero}
\eeq
for any $A\in\mfa(\mco)$ and $f$ as defined above. Since equality (\ref{zero}) extends to any $A\in\mfa$,
we conclude that $\om$ is a vacuum state in the sense of Definition 4.3 from \cite{Ara}. Invariance of $\om$
under translations and validity of the relativistic spectrum condition in its GNS-representation follow from
 Theorem 4.5 of \cite{Ara}, provided that the functions $\real^{s+1}\ni x\to\om(A^*B(x))$ are continuous for any $A, B\in \mfa$. Since local operators form a norm-dense subspace of $\mfa$, it is enough to prove continuity for 
$A,B\in\mfa(\mco)$ for any open, bounded region $\mco$. For this purpose we recall from \cite{BP} 
that Condition~$\Cs$ has a dual formulation which says that the maps $\Xi_E:\mfa(\mco)\to B(\hil)$ given
by $\Xi_E(A)=P_EAP_E$ are compact for any open, bounded region $\mco$ and any $E\geq 0$. Given any sequence
of spacetime points $x_n\to x$, there holds $A^*(B(x_n)-B(x))\to 0$ in the strong topology and, by compactness of the maps $\Xi_E$, $P_E A^*(B(x_n)-B(x))P_E\to 0$ in the norm topology in $B(\hil)$. Now the required continuity follows from the bound
\beq
|\om\big(A^*(B(x_n)-B(x))\big)|\leq \|P_E A^*(B(x_n)-B(x))P_E\|
\label{strongcont}
\eeq
which can be established with the help of the approximating sequence $\fun_n\in\traceEB$.\\
(b) We note that for any open, bounded region $\mco$, $E\geq 0$ and  $\eps>0$, Condition $\Nnat$ allows for such $N$ and $\x$, belonging to the spacelike ray,
that $2N^{-\h}\|\PiE\|_{\x}\leq \fr{\eps}{3}$. For arbitrary $A\in\mfa(\mco)_1$ we can find $\fun\in\traceEB$ s.t. 
$\sup_{k\in\{1,\ldots,N\}}|\om(A(x_k))-\fun(A(x_k))|\leq\fr{\eps}{3}$ and $|1-\fun(I)|\leq\fr{\eps}{3}$. Next, we note that
\beqa
& &|\om(A)-\om_0(A)|\leq|\om(A)-\fun(I)\om_0(A)|+\fr{\eps}{3}\non\\
& &\phantom{44444}\leq\fr{1}{N}\sum_{k=1}^N\big|\al_{x_k}^*\om(A)-\al^*_{x_k}\fun(A)\big|
+\fr{1}{N}\sum_{k=1}^N \big|\al^*_{x_k}\fun(A)-\fun(I)\al^*_{x_k}\om_0(A)\big|+\fr{\eps}{3}\non\\
& &\phantom{44444}\leq\sup_{k\in\{1,\ldots, N\}}
|\om(A(x_k))-\fun(A(x_k))|+2N^{-\h}\|\PiE\|_{\x}+\fr{\eps}{3}\leq\eps,
\eeqa
where in the second step we made use of the fact that both $\om$ and $\om_0$ are invariant under the
translations $\x$ and in the third step we used the H\"older inequality and the fact that $\h(\fun-\fun(I)\om_0)\in\tracB$. We conclude that the states $\om$ and $\om_0$ coincide on any local operator and therefore on the whole algebra $\mfa$. \qed\\
The above result is of relevance to the problem of convergence of physical states to the vacuum under large timelike translations. In fact, the following lemma asserts that the respective limit points are invariant under translations in some spacelike hyperplane.
\begin{lemma}[D.Buchholz, private communication]\label{time} Suppose that Condition $\Cs$ holds. Let $\omex$ be a weak* limit point as $\lam\to\infty$ of the net $\{\al_{\lam \e}^*\om\}_{\lam\in\real_+}$ of states on $\mfa$,
where $\e\in\real^{s+1}$ is a timelike unit vector and $\om$ is a state from $\traceE$ for some $E\geq 0$.
Then $\omex$ is invariant under translations in the spacelike hyperplane 
$\{\e^\perp\}=\{  x\in\real^{s+1} \ | \ \e\cdot x =0  \}$, where dot denotes the Minkowski scalar product.
\eel
\proof
Choose $x\in\{\e^\perp\}$, $x\neq 0$. Then there exists a Lorentz transformation $\La$ and 
$y^0,y^1\in\real\backslash\{0\}$ s.t.
$\La \e=y^0\e_0$, $\La x = y^1\e_1$, where $\e_{\mu}$, $\mu=0,1,\ldots,s$ form the canonical basis in $\real^{s+1}$. We set $v=\fr{y^1}{y^0}$ and introduce the family of Lorentz transformations $\La_t=\La^{-1}\tilde{\La}_t\La$, where $\tilde{\La}_t$ denotes the boost in the direction of $\e_1$ with rapidity $\textrm{arsinh}(\fr{v}{t})$.
By the composition law of the Poincar\'e group, the above transformations composed with
translations in timelike direction give also rise to spacelike translations
\beq
(0,\La_\lam)(\lam \e,I)(0,\La_\lam^{-1})=(\lam\La_\lam \e,I),\quad
\lam\La_\lam \e=\lam\sqrt{1+\big(v/\lam\big)^2}\e+x.
\eeq
We make use of this fact in the following estimate: 
\beqa
|\al_{\lam \e}^*\om(A)-\al_{\lam \e}^*\om(A(x))|&\leq&
|\om(\al_{\lam \e}A)-\om(\al_{\La_{\lam}}\al_{\lam \e}\al_{\La_{\lam}^{-1}}A)|
\nonumber\\
&+&|\al_{\lam\La_\lam \e}^*\om(A)-\al_{\lam \e}^*\om(A(x))|,
\label{translate}
\eeqa
where $A\in\mfa(\mco)$. The first term on the r.h.s. of (\ref{translate}) satisfies the bound
\beqa
& &|\om(\al_{\lam \e}A)-\om\big(\al_{\La_{\lam} }\al_{\lam \e}\al_{\La^{-1}_{\lam} }A\big)|\non\\
& &\phantom{444444444}\leq|\al_{\lam \e}^*\om(A-\al_{\La^{-1}_{\lam}}A)|+|(\om-\al_{\La_{\lam}}^*\om)
(\al_{\lam \e}\al_{\La_{\lam }^{-1}}A)|\non\\
& &\phantom{444444444}\leq \|P_{E}(A-\al_{\La_{\lam }^{-1}}A)P_{E}\|+
\sup_{s\in\real_+}\|\om-\al_{\La_{\lam }}^*\om \|_{\mfa(\widetilde{\mco}+s\e)}  \|A\|,
\eeqa
where $\widetilde{\mco}$ is a slightly larger region than $\mco$.
Clearly, $\La_{\lam}\to I$ for $\lam\to\infty$ and  therefore $\al_{\La_{\lam}}\to \textrm{id}$ in the point - weak open topology. Then the above expression tends to zero in this limit by the dual form of Condition $\Cs$ and the assumption that Lorentz transformations are unitarily implemented. (The argument is very similar to the last step in the proof of Theorem \ref{disconnected}~(a). We note that the restriction on Lorentz transformations can be relaxed to a suitable regularity condition). 
The second term on the r.h.s. of (\ref{translate}) converges to zero by the dual variant of Condition~$\Cs$ 
and the following bound:
\beqa
& &|\al_{\lam\La_\lam \e}^*\om(A)-\al_{\lam \e}^*\om(A(x))|\leq
|\om\big(A\big(\lam\sqrt{1+\big(v/\lam\big)^2 }\e+x\big)-A(\lam \e+x)\big)|\non\\
& &\phantom{44444444444444} \leq\|P_{E}\big(A\big(\big\{\sqrt{1+\big(v/\lam\big)^2}+1\big\}^{-1}(v^2/\lam)  \e\big)-A\big)P_{E}\|.
\eeqa  
Thus we demonstrated that  $\omex(A)=\omex(A(x))$ for any local operator $A$.
This result extends by continuity to any $A\in\mfa$. \qed\\ 
It follows from Theorem~\ref{disconnected} (a) that all the limit points $\omex$ are vacuum states under 
the premises of the above lemma. On the other hand, adopting Condition~$\Nnat$ we obtain a stronger result from 
Theorem~\ref{disconnected} (b):
\bec Let Condition $\Nnat$ be satisfied. Then, for any state $\om\in\traceE$, $E\geq 0$, and 
timelike unit vector $\e\in\real^{s+1}$, there holds
\beq
\lim_{t\to\infty} \al_{t\e}^*\om(A)=\om_0(A),\textrm{ for } A\in\mfa.
\eeq
\eec
\noindent We note that in contrast to previous approaches to the problem of relaxation to the vacuum \cite{AH,BWa} the present argument does not require the assumption of asymptotic completeness or asymptotic abelianess in time.

To conclude this survey of applications of Condition $\Nnat$ let us mention 
another physically meaningful procedure for preparation of vacuum states: It is to construct states with 
increasingly sharp values of energy and momentum and exploit the uncertainty principle.
Let $\Ppr$ be the spectral projection corresponding to the ball of radius $r$ centered around point $p$ in the
energy-momentum spectrum. Then, in a theory satisfying Condition $\Nnat$,  any sequence of states $\om_r\in\Ppr\trace\Ppr$ converges, uniformly on local algebras, to the vacuum state $\om_0$ as $r\to 0$, since this is the only energetically
accessible state which is completely dislocalized in spacetime. This fact is reflected in the following property of the map $\PiE$: 
\bep\label{shrinking} Suppose that Condition $\Nnat$ is satisfied. Then, for any $E\geq0$ and $p\in\cone$, there holds
\beq
\lim_{r\to 0}\|\PiE|_{\traceprc} \|=0,
\eeq
where $\traceprc=\{\fun-\fun(I)\om_0 \ | \ \fun\in \Ppr\traceE\Ppr \}$.
\eep
\proof 
We pick $A\in B(\hil)$, $\fun\in\traceprc$ and estimate the deviation of this functional 
from translational invariance
\beqa
|\fun(A)-\al^*_x\fun(A)|&=&|\fun(\Ppr A \Ppr)-\fun(\Ppr e^{i(P-p)x}A e^{-i(P-p)x}\Ppr)|\non\\
&=&|\fun(\Ppr e^{i(P-p)x}A(1-e^{-i(P-p)x})\Ppr)\non\\ 
&+&\fun(\Ppr(1-e^{i(P-p)x})A\Ppr)|\leq 2\|\fun\|\,\|A\|\,|x|\,\rad,
\eeqa
where in the first step we used invariance of $\om_0$ under translations to insert the
projections $P_{(p,r)}$ and in the last step we applied the spectral theorem. Consequently, for any
$x_1,\ldots, x_N\in\real^{s+1}$ and open bounded region $\mco$
\beqa
\|\fun\|_{\mfa(\mco)}&\leq& \fr{1}{N}\sum_{k=1}^N\|\al^*_{x_k}\fun\|_{\mfa(\mco)}
+\sup_{k\in\{1,\ldots,N\}}\|\fun-\al^*_{x_k}\fun\|_{\mfa(\mco)}\non\\
&\leq&\fr{1}{\sqrt{N}}\big(\sum_{k=1}^N\|\al^*_{x_k}\fun\|_{\mfa(\mco)}^2\big)^\h+
 2\|\fun\|\,\rad\sup_{k\in\{1,\ldots,N\}} |x_k|.
\eeqa
To conclude the proof of the proposition we restate the above inequality as follows:
\beq
\|\PiE|_{\traceprc}\|\leq\fr{1}{\sqrt{N}}\|\PiE\|_{x_1,\ldots,x_N}+2\rad\sup_{k\in\{1,\ldots,N\}} |x_k|,
\eeq
and make use of Condition $\Nnat$. \qed\\
It is a consequence of the above proposition that $\lim_{E\searrow 0}\N(\eps)_{E}=1$ 
in any theory complying with Condition $\Nnat$, as anticipated in our heuristic discussion.
Since $\N(\eps)_{E}\geq 1$ and it decreases monotonically with decreasing $E$, the limit exists.
If it was strictly larger than one,  we could find nets of functionals $\fun_{1,E},\fun_{2,E}\in \tracB$
s.t. $\|\Pi_E(\fun_{1,E}-\fun_{2,E})\|>\eps$ for any $E>0$. But fixing some $E_0>0$ and restricting
attention to $E\leq E_0/\sqrt{2}$ we obtain
\beq
\eps<\|\Pi_E(\fun_{1,E}-\fun_{2,E})\|\leq 2\|\Pi_{E_0}|_{\traceprcz}\|.
\eeq
The last expression on the r.h.s. tends to zero with $E\to 0$, by Proposition \ref{shrinking},
leading to a contradiction.

Up to this point we discussed the physical interpretation and applications of the novel
Condition $\Nnat$ from the general perspective of local relativistic quantum field theory.
In order to shed more light on the mechanism which enforces this and related phase space criteria, 
we turn now to their verification in a model.

\section{Condition $\Ns$ in Massive Scalar Free Field Theory}
\setcounter{equation}{0}
In this section, which serves mostly to fix our notation, we recall some basic 
properties of scalar free field theory of mass $m>0$ in $s$ space dimensions. (See \cite{RS2} Sect. X.7).
The single particle space of this theory is $L^2(\real^s, d^sp)$. On this space there act the multiplication operators $\om(\vep)=\sqrt{|\vep|^2+m^2}$ and $p_1, \ldots, p_s$ which are self-adjoint on a suitable dense domain
and generate the  unitary representation of translations
\beq 
(U_1(x)f)(\vep)=e^{i(\om(\vep)x^0-\vep \vx)}f(\vep),\quad f\in L^2(\real^s, d^sp).
\eeq
The full  Hilbert space $\hil$ of the theory is the symmetric Fock space over
$L^2(\real^s, d^sp)$. By the method of second quantization we obtain 
the Hamiltonian $H=d\Ga(\om)$, and the momentum operators $P_i=d\Ga(p_i)$, $i=1,2,\ldots,s$ 
defined on a suitable domain in $\hil$. The joint spectrum of this family of commuting, self adjoint 
operators is contained in the closed forward light cone. The unitary representation
of translations in $\hil$  given by
\beq
U(x)=\Ga(U_1(x))=e^{i(Hx^0-\veP\vx)}
\eeq
implements the corresponding family of automorphisms of $B(\hil)$
\beq
\al_x(\cdot)=U(x)\cdot U(x)^*.
\eeq
Next, we construct the local algebra $\mfa(\mco)$ attached to the double cone $\mco$,
whose base is the $s$-dimensional ball $\mco_r$ of radius $r$ centered at the origin in
configuration space. To this end we introduce the subspaces $\Lpm=[\om^{\mp\fr{1}{2}}\widetilde{D}(\mco_r)]$, where
tilde denotes the Fourier transform. (The respective projections are denoted by $\Lpm$ as well.)
Defining $J$ to be the complex conjugation in  configuration space
we introduce the real linear subspace
\beq
\LJ=(1+J)\Lp+(1-J)\Lm
\eeq
and the corresponding von Neumann algebra
\beqa
\mfa(\mco)=\{ \ W(f) \ | \ f\in\LJ \}^{\prime\prime},
\eeqa
where $W(f)=e^{i(a^*(f)+a(f))}$ and $a^*(f)$, $a(f)$ are the creation and annihilation operators. 
With the help of the translation automorphisms $\al_x$ introduced above we define local 
algebras attached to double cones centered at any point $x$ of spacetime
\beq
\mfa(\mco+x)=\al_x(\mfa(\mco)).
\eeq
The global algebra $\mfa$ is the $C^*$-inductive limit of all such local algebras
of different $r>0$ and $x\in\real^{s+1}$. By construction, $\al_x$ leaves $\mfa$ invariant.

Now we turn our attention to the phase space structure of the theory. 
Let $Q_E$ be the projection on states of energy lower than $E$ in the single particle space 
and $\be\in\real$.
We define operators $\SEpm=Q_E\Lpm$, $\Sbpm=e^{-\half(\be|\vep|)^2}\Lpm$.
It  follows immediately from \cite{modular}, p.~137 that these operators satisfy
$\||\SEpm|^p\|_1<\infty$, $\||\Sbpm|^p\|_1<\infty$ for any $p>0$, where $\|\cdot\|_1$ denotes
the trace norm. We introduce their least upper bound $\Sa$
\beq
\Sa=\textrm{s-}\lim_{n\to\infty}\bigg(\fr{1}{4}(|\SEp|^{2^n}+|\SEm|^{2^n}+|\Sbp|^{2^n}+|\Sbm|^{2^n})\bigg)^{2^{-n}}.
\label{operatorT}
\eeq
Proceeding as in \cite{BJ1} p. 316/317 one can show that this limit exists and that the operator $\Sa$ satisfies
\beqa
 \Sa^n &\geq& |\SEpm|^n \textrm{ and } \Sa^n\geq |\Sbpm|^n \textrm{ for } n\in\nat,\\
\|\Sa\|&\leq& \max(\|\SEp\|,\|\SEm\|,\|\Sbp\|,\|\Sbm\|)\leq 1,\label{lub2}\\
\|\Sa^p\|_1 &\leq& \||\SEp|^p\|_1+\||\SEm|^p\|_1 +\||\Sbp|^p\|_1+\||\Sbm|^p\|_1 \textrm{ for } p>0. \label{lub3}
\eeqa
In particular $\Sa$ is a trace class operator. Since it commutes with the conjugation $J$, 
the orthonormal basis of its eigenvectors $\{e_j\}_1^\infty$ 
can be chosen so that $Je_j=e_j$. The corresponding eigenvalues will be denoted $\{\s_j\}_1^\infty$. 
Given any pair of multiindices $\mub=(\mup,\mum)$ we define the operator
\beq
\Bm=a(\LL e)^{\mub}=a(\Lp e)^{\mup}a(\Lm e)^{\mum}.
\eeq
We recall, that for any $f_1,\ldots,f_n\in L^2(\real^s, d^sp)$ there hold the so called energy bounds \cite{BP}
which in the massive theory have the form
\beq
\|a(f_1)\ldots a(f_n)P_E\|=\|P_E a^*(f_n)\ldots a^*(f_1)\|\leq (\M)^{\fr{n}{2}}\|f_1\|\ldots \|f_n\|,
\eeq
where $\M=\fr{E}{m}$.
Consequently, the operators $\Bm$ are bounded on states of finite energy. We note the respective bound
\beqa
\|\Bm P_E\|\leq \|a(Q_E\LL e)^{\mub}P_E\|&\leq& (\M)^{\fr{|\mub|}{2}}
\| Q_E\LL e\|^{\mub}\non\\
&\leq& (\M)^{\fr{|\mub|}{2}}\s^{\mub},\label{Bnorm}
\eeqa
where $|\mub|=|\mup|+|\mum|$, $\s^{\mub}=\s^{\mup}\s^{\mum}$, $\{\s_j\}_1^\infty$ are the eigenvalues of $\Sa$ 
and in the last step we made use of the fact that $|Q_E\Lpm|^2\leq \Sa^2$.
We will construct the expansion of $\PiE$ into rank-one maps with the help of the 
bounded linear functionals ${\T}_{\mub,\nub}: \trac\to\complex$, given by
\beq
\Tmn(\fun)=\fun(\Bm^*\Bn).
\eeq
In particular $\T_{0,0}=0$, since $\fun(I)=0$ for any $\fun\in\trac$.
It follows from (\ref{Bnorm}) that the norms of these 
maps satisfy the bound
\beq
\|\Tmn\|\leq \M^{\fr{|\mub|+|\nub|}{2}}\s^{\mub} \s^{\nub}.
\eeq 
Clearly, we can assume that $\M\geq 1$ as $\PiE\equiv 0$ otherwise.
Since  $\Tmn=0$ for $|\mub|>\M$ or $|\nub|>\M$,  the norms of the functionals $\Tmn$ are summable with any power 
$p>0$. In fact
\beqa
\sum_{\mub,\nub}\|\Tmn\|^p &\leq& \M^{p\M}(\sum_{\mub:|\mub|\leq \M}\s^{p\mub})^2
\leq \M^{p\M} (\sum_{\mup:|\mup|\leq \M}\s^{p\mup})^4\non\\
&=&\M^{p\M} (\sum_{k=0}^{[\M]}\sum_{\mup:|\mup|=k}\s^{p\mup})^4
\leq \M^{p\M} (\sum_{k=0}^{[\M]}\|\Sa^p\|_1^k)^4, \label{traces}
\eeqa
where in the last step we made use of the multinomial formula.
With this information at hand it is easy to verify that Condition $\Ns$ holds in massive
scalar free field theory \cite{BP,Bos3}.
\bet\label{sharp}
In massive scalar free field theory there exist functionals $\tau_{\mub,\nub}\in\mfa(\mco)^*$
such that there holds in the sense of norm convergence in $\mfa(\mco)^*$
\beq
\PiE(\fun)=\sum_{\mub,\nub}\tau_{\mub,\nub} \T_{\mub,\nub}(\fun), \quad \fun\in\trac. \label{ffexpansion}
\eeq
Moreover, $\|\tau_{\mub,\nub}\|\leq 2^{5\M}$ for all $\mub,\nub$ and
 $\sum_{\mub,\nub} \|\T_{\mub,\nub}\|^p<\infty$ for any $p>0$.
\eet
\noindent We give the proof of this theorem in  Appendix B.
\section{Condition $\Nnat$  in Massive Scalar Free Field Theory}
\setcounter{equation}{0}
At this point we turn to the main goal of this technical part of our investigations, namely to
verification of Condition $\Nnat$ in the model at hand.
By definition of the nuclear $p$-norms and Theorem \ref{sharp} there holds the bound
\beq
\|\PiE\|_{p,\x}\leq\bigg(\sum_{\mub,\nub}\|\tau_{\mub,\nub}\|^p\|\T_{\mub,\nub}\|^p_{\x}\bigg)^{\pin}
\leq 2^{5\M}\bigg(\sum_{\mub,\nub}\|\T_{\mub,\nub}\|^p_{\x}\bigg)^{\pin}. \label{start}
\eeq
Consequently, we need estimates on the norms $\|\Tmn\|_{\x}$ whose growth with $N$
can be compensated by  large spacelike distances $x_i-x_j$ 
for $i\neq j$. This task will be accomplished in Proposition \ref{semibound}.
The argument is based on the following lemma which is a variant of Lemma 2.2 from \cite{Buch3}.
\bel\label{harmonic} Let $B$ be a (possibly unbounded) operator s.t. $\|BP_E\|<\infty$, $\|B^*P_E\|<\infty$
and $BP_E\hil\subset P_{E-m}\hil$ for any $E\geq 0$. Then, for any  $x_1,\ldots,x_N\in\real^{s+1}$, there
hold the bounds
\begin{enumerate}
\item[(a)] $ \|P_E\sum_{k=1}^N(B^*B)(x_k)P_E\|\leq (\M+1)\bigg\{\|P_E[B,B^*]P_E\|\non\\
\phantom{4444444444444444444444}+(N-1)\sup_{k_1\neq k_2}\|P_E[B(x_{k_1}),B^*(x_{k_2})]P_E\|\bigg\},$
\item[(b)] $ \|P_E\int_{K}d^sx(B^*B)(\vx)P_E\|\leq (\M+1)\int_{\De K} d^sx\|P_E[B(\vx),B^*]P_E\|,$
\end{enumerate}
where $K$ is a compact subset of $\real^s$ and $\De K=\{\vx-\vy \ | \ \vx,\vy\in K \}$.
\eel
\proof Part (b) coincides, up to minor modifications, with \cite{Buch3}.
In the proof of part~(a) the modifications are more substantial, so we provide
some details. We will show, by induction in $n$, that there holds the following inequality:
\beqa
& &\|P_{nm}\sum_{k=1}^N(B^*B)(x_k)P_{nm}\|
\leq n\bigg\{\|P_{(n-1)m}[B,B^*]P_{(n-1)m}\|\non\\
& &\phantom{44444444444}+(N-1)\sup_{k_1\neq k_2}\|P_{(n-1)m}[B(x_{k_1}),B^*(x_{k_2})]P_{(n-1)m}\|\bigg\}, \label{discrete}
\eeqa
where $P_{nm}$ is the spectral projection of $H$ on the subspace spanned by vectors of energy lower than $nm$.
It clearly holds for $n=0$. To make the inductive step we pick 
$\om(\bdot)=(\Phi|\bdot|\Phi)$, $\Phi\in(P_{nm}\hil)_1$ and define
$Q=\sum_{k=1}^N(B^*B)(x_k)$. Proceeding like in \cite{Buch3}, with integrals replaced
with sums, one arrives at
\beqa
\om(QQ)&\leq&\sum_{k=1}^N\om((B^*B)(\vx_k))\big\{\sum_{l=1}^N \|P_{(n-1)m}[B(\vx_l),B^*(\vx_k)]P_{(n-1)m}\|\big\}\non\\
&+&\om(Q)\|P_{(n-1)m}QP_{(n-1)m}\|. \label{QQ}
\eeqa
The sum w.r.t. $l$ in the first term on the r.h.s. can be estimated by the expression in 
curly brackets in (\ref{discrete}). To the second term on the r.h.s. of (\ref{QQ}) we apply the
induction hypothesis. Altogether
\beqa
\om(QQ)&\leq& n\om(Q)\bigg\{\|P_{(n-1)m}[B,B^*]P_{(n-1)m}\|\non\\
&+&(N-1)\sup_{k_1\neq k_2}\|P_{(n-1)m}[B(x_{k_1}),B^*(x_{k_2})]P_{(n-1)m}\|\bigg\}.
\eeqa
Making use of the fact that $\om(Q)^2\leq\om(QQ)$ and taking the supremum over states $\om$ which 
are induced by vectors from $P_{nm}\hil$ one concludes the proof of  estimate~(\ref{discrete}). 
The statement of the lemma follows by choosing $n$ s.t. $(n-1)m\leq E \leq nm$. 
\qed\\
In order to control the commutators appearing in the estimates in Lemma \ref{harmonic}  
we need a slight generalization of the result from \cite{Fredenhagen} on the exponential decay of vacuum 
correlations between local observables.
\bet\label{damping} Let $H$ be a self-adjoint operator on a Hilbert space
$\hil$ s.t. $Sp H=\{0\}\cup [m,\infty]$, $m>0$ and there exists exactly one
(up to a phase) eigenvector $\vac$ of $H$ with eigenvalue zero. Let $A$, $B$
be operators such that $\vac$ belongs to their domains and to the domains of their adjoints.
If there holds
\beq
(\vac| \, [A, e^{itH}Be^{-itH}] \, \vac)=0 \textrm{ for } |t|<\de,
\eeq
then
\beq
|(\vac|AB\vac)-(\vac|A\vac)(\vac|B\vac)|\leq e^{-m\de}
\{\|A\vac\| \, \|A^*\vac\| \, \|B\vac\| \, \|B^*\vac\|\}^{\half}.
\eeq
\eet
\noindent With the help of the above theorem we prove the desired estimate.
\bel\label{bounds} Let $e\in L^2(\real^s, d^sp)$ be s.t. $\|e\|\leq 1$ and $Je=e$. Then there holds
for any $x\in\real^{s+1}$, $0<\eps<1$ and any combination of $\pm$ signs
\beq
|\lan \Lpm e|e^{-(\be|\vep|)^2} U(x)\Lpm e \ran|\leq c_{\eps,\be}e^{-m(1-\eps)\de(x)}, \label{gaussian}
\eeq
where $c_{\eps,\be}$ does not depend on $x$ and $e$. Here $\de(x)=|\vx|-|x^0|-2r$ and $r$ is the 
radius of the double cone entering into the definition of the projections $\Lpm$.
\eel
\proof 
We define the operators $\phi_{+}(e)=a^*(\Lp e)+a(\Lp e)$, $\phi_{-}(e)=a^*(i\Lm e)+a(i\Lm e)$ and their
translates $\phi_{\pm}(e)(x)=U(x)\phi_{\pm}(e)U(x)^{-1}$. Since the projections $\Lpm$ commute with
$J$ and $Je=e$, these operators are just the 
fields and canonical momenta of massive scalar free field theory. Assume that  $\de(x)>0$. Then, by locality,
$\phi_{\pm}(e)$ and $\phi_{\pm}(e)(x)$ satisfy the assumptions of Theorem \ref{damping}.
As  they have vanishing vacuum expectation values, we obtain 
\beq
|\lan \Lpm e|U(x)\Lpm e\ran|=|(\vac|\phi_{\pm}(e)\phi_{\pm}(e)(x)\vac)|
\leq e^{-m\de(x)}. \label{nogaussian}
\eeq
Let us now consider the expectation value from the statement of the lemma. We fix some $0<\eps<1$
and estimate
\beqa
& &|\lan \Lpm e|e^{-(\be|\vep|)^2}U(x)\Lpm e\ran|\non\\
& &\phantom{444444444}\leq 
\cbe\int_{\de(\vy+x)\geq (1-\eps)\de(x)} d^sy \ 
e^{-\fr{|\vy|^2}{4\be^2}}|\lan \Lpm e|U(x+\vy)\Lpm e\ran|\non\\
& &\phantom{444444444}+\cbe\int_{\de(\vy+x)\leq (1-\eps)\de(x)} d^sy \ e^{-\fr{|\vy|^2}{4\be^2}}|\lan \Lpm e|U(x+\vy)\Lpm e\ran|\non\\
& &\phantom{444444444}\leq e^{-m(1-\eps)\de(x)}+\cbe\int_{|\vy|\geq \eps\de(x)}d^sy \ e^{-\fr{|\vy|^2}{4\be^2}}\non\\
& &\phantom{444444444}\leq e^{-m(1-\eps)\de(x)}\bigg(1+\cbe\int d^sy \ e^{-\fr{|\vy|^2}{4\be^2}+\fr{m(1-\eps)|\vy|}{\eps} }\bigg). \label{division}
\eeqa
In the first step we expressed the function $e^{-(\be|\vep|)^2}$ by its  Fourier transform and divided the region 
of integration into two subregions. To the first integral we applied estimate
(\ref{nogaussian}). Making use of the fact that the second integral decays faster than exponentially with 
$\de(x)\to\infty$, we arrived at the last expression which is of the form (\ref{gaussian}). Since
$c_{\eps,\be}>1$, the bound (\ref{division}) holds also for  $\de(x)\leq 0$. \qed

It is a well known fact that any normal, self-adjoint functional on a von Neumann algebra can be
expressed as a difference of two normal, positive functionals which are mutually orthogonal \cite{Sakai}.
It follows that any $\fun\in\traceEB$ can be decomposed as 
\beq
\fun=\funr^+-\funr^-+i(\funi^+-\funi^-), \label{decomp}
\eeq
where $\funr^\pm$, $\funi^\pm$ are positive functionals from $\traceEB$. This assertion
completes the list of auxiliary results needed to establish the required estimate for $\|\Tmn\|_{\x}$.
\bep\label{semibound} The functionals $\Tmn$   satisfy the bound
\beq
\|\Tmn\|_{\x}^2\leq 
32 \s^{\mub} \s^{\nub} (\M)^{2\M}e^{(\be E)^2}\big\{1+\sqrt{c_{\eps,\be} }(N-1)e^{-\fr{m}{2}(1-\eps)\de(\vxxb)}  \big\},
\eeq
where $\{\s_j\}_1^{\infty}$ are the eigenvalues of the operator 
$\Sa$ given by formula (\ref{operatorT}) and $\de(\vxxb)=\inf_{i\neq j} \de(x_i-x_j)$. The function $\de(x)$, the parameter $\eps$ and the constant $c_{\eps,\be}$ appeared in Lemma \ref{bounds}.
\eep
\proof 
We denote by $\traceEBP$ the set of positive functionals from $\traceEB$. Making use 
of the definition of $\|\cdot\|_{\x}$, decomposition (\ref{decomp}) and the Cauchy-Schwarz inequality
we obtain
\beqa
\|\T_{\mub,\nub}\|^2_{\x}&=&\sup_{\fun\in\tracB}\sum_{k=1}^N|\Tmn(\al_{x_k}^\p\fun)|^2 
 \leq 16\sup_{\fun\in\traceEBP}\sum_{k=1}^N|\al_{x_k}^\p\fun(\Bm^*\Bn) |^2 \non\\
& &\phantom{4444444444}\leq 16\sup_{\fun\in\traceEBP}\sum_{k=1}^N\al_{x_k}^\p\fun(\Bm^*\Bm)\al_{x_k}^\p\fun(\Bn^*\Bn)\non\\
& &\phantom{4444444444}\leq 16(\M)^{|\mub|}\s^{2\mub}\| P_E\sum_{k=1}^N(\Bn^*\Bn)(x_k)P_E\|,\qquad \label{seminorm1}
\eeqa
where in the last step we
applied the bound (\ref{Bnorm}). We can assume, without loss of generality, that $\nub\neq 0$
and decompose it into two pairs of multiindices $\nub=\nuh+\nuc$ in such a way that $|\nuc|=1$. Since $\Bn=\Bnh\Bnc$, we get
\beqa
P_E\sum_{k=1}^N(\Bn^*\Bn)(x_k)P_E&=&P_E\sum_{k=1}^N(\Bnc^*P_E\Bnh^*\Bnh P_E\Bnc)(x_k)P_E\non\\
&\leq&\|\Bnh P_E\|^2 P_E\sum_{k=1}^N(\Bnc^*\Bnc)(x_k)P_E\non\\
&=&\M^{|\nuh|}\s^{2\nuh}  P_E\sum_{k=1}^N\big(a^*(\LL e)^{\nuc}a(\LL e)^{\nuc}\big)(x_k)P_E,\qquad
\label{aux2}
\eeqa
where in the last step  we used again estimate (\ref{Bnorm}).
Next, let $g$ be the operator of multiplication by $\half(\be|\vep|)^2$ in $L^2(\real^s,d^sp)$
and let $G=d\Ga(g)\geq 0$  be its second quantization.
Since one knows explicitly the action of $G$ and $H$ on vectors of fixed particle number,
it is easy to check that
\beq
e^G P_E=P_E e^G P_E\leq P_Ee^{\half (\be H)^2}P_E\leq e^{\half (\be E)^2}.
\eeq
Making use of this fact, Lemma \ref{harmonic} (a) and Lemma \ref{bounds} we obtain from (\ref{aux2})
the following string of inequalities:
\beqa
& &\|P_E\sum_{k=1}^N(\Bn^*\Bn)(x_k)P_E\|\non\\
& &\phantom{4444} \leq\M^{|\nuh|}\s^{2\nuh} 
\|P_Ee^G\sum_{k=1}^N\big(a^*(\damp\LL e)^{\nuc}e^{-2G}a(\damp\LL e)^{\nuc}\big)(x_k)e^GP_E\|\non\\
& &\phantom{4444}\leq\M^{|\nuh|}\s^{2\nuh} 
e^{(\be E)^2}\|P_E\sum_{k=1}^N\big(a^*(\damp\LL e)^{\nuc}a(\damp\LL e)^{\nuc}\big)(x_k)P_E\|\non\\
& &\phantom{4444}\leq\M^{|\nuh|}\s^{2\nuh} 
e^{(\be E)^2}(\M+1)\big\{\lan(\LL e)^{\nuc}|\dam(\LL e)^{\nuc}\ran\non\\
& &\phantom{44444444444444444444}+(N-1)\sup_{i\neq j}|\lan(\LL e)^{\nuc}|\dam U(x_i-x_j)(\LL e)^{\nuc}\ran| \big\}\non\\
& &\phantom{4444}\leq 2\M^{|\nub|}\s^{\nub}
e^{(\be E)^2}\big\{1+(N-1)\sqrt{c_{\eps,\be}}\sup_{i\neq j} e^{-\fr{m}{2}(1-\eps)\de(x_i-x_j)} \big\}, \label{collect}
\eeqa
where in the last step we made use of the estimate $\phantom{44}
|\lan \Lpm e_j |\dam U(x)\Lpm e_j\ran|\leq \\ \lan e_j | |\Sbpm|^2 e_j\ran \leq \lan e_j | \Sa^2 e_j\ran=\s_j^2$ and  the fact that $\s_j\leq 1$ which follows from (\ref{lub2}). Substituting inequality
(\ref{collect}) to  formula (\ref{seminorm1}), estimating  $\s^{2\mub}\leq \s^{\mub}$
and recalling that $\Tmn=0$ for $|\mub|>\M$ or $|\nub|>\M$ we obtain the bound from the 
statement of the proposition. \qed\\
It is now straightforward to estimate the  $p$-norms of the map $\PiE$. Substituting the bound from the
above proposition to formula (\ref{start}) and proceeding like in estimate~(\ref{traces}) we obtain
\beqa
& &\|\PiE\|_{p,\x}\non\\ 
& &\leq (4\sqrt{2}) (2^5\M)^{\M}e^{\half(\be E)^2}
\big(\sum_{k=0}^{[\M]}\|\Sa^\fr{p}{2}\|_1^k\big)^{\fr{4}{p}}
\big\{1+\sqrt{c_{\eps,\be} }(N-1)e^{-\fr{m}{2}(1-\eps)\de(\vxxb)}  \big\}^{\fr{1}{2}}.\,\,\qquad
\eeqa
It is clear from the above relation that  $\limsup_{\de(\vxxb)\to\infty}\|\PiE\|_{p,\x}$ satisfies a bound which is
independent of $N$.  Consequently, we get
\bet Condition $\Nnat$ holds in massive scalar free field theory for arbitrary dimension of space $s$.
\eet

\section{Conclusion and Outlook}
\setcounter{equation}{0}
In this work we proposed and verified in massive scalar free field theory the new  Condition $\Nnat$.
Since this phase space criterion encodes the  firm physical principle that energy is additive over isolated subsystems, we expect that it holds in a large family of models. In fact,
we will show in a future publication that massless scalar free field theory also satisfies this condition for 
$s\geq 3$. We recall that this model contains an infinite family of pure, regular vacuum states which are, however, mutually energy-disconnected \cite{BWa}. In view of Theorem \ref{disconnected} (b), this decent vacuum structure is related to phase space properties of this model, as anticipated in \cite{BP}.

Apart from more detailed information about the phase space structure of 
massive free field theory, our discussion offers also some new insights 
into the harmonic analysis of translation automorphisms. First, we recall from 
\cite{Buch3} that in all local, relativistic quantum field theories 
there holds the bound
\beq
\sup_{\fun\in\traceEB}\int d^sp|\vep|^{s+1+\eps}|\fun(\widetilde{A}(\vep))|^2<\infty, \label{ha}
\eeq
for any $\eps>0$, uniformly in $A\in\mfa(\mco)_1$. It says that the distribution $\fun(\widetilde{A}(\vep))$, restricted
to the domain $\{\vep \ | \ |\vep|\geq\de  \}$ for some $\de>0$, is represented by a 
square integrable function, but at $\vep=0$ it may have a power like singularity which is not square
integrable. It turns out, however, that in massive scalar free field theory this distribution has a milder behavior
at zero than one might expect from (\ref{ha}). 
Making use of Lemma \ref{harmonic} (b) and going through our argument once again,
one can easily establish that there holds, uniformly in $A\in\mfa(\mco)_1$,
\beq
\sup_{\fun\in\traceEB}\int d^sx|\fun(\oA(\vx))|^2<\infty,
\eeq
where $\oA=A-\om_0(A)I$. By the Plancherel theorem, we obtain
\beq
\sup_{\fun\in\traceEB}\int d^sp|\fun(\widetilde{\oA}(\vep))|^2<\infty,
\eeq
i.e. the distribution $\fun(\widetilde{\oA}(\vep))$ is represented by a square integrable 
function. Consequently, $\fun(\widetilde{A}(\vep))$ can deviate from square integrability only 
by a delta-like singularity at $\vep=0$.
The above reasoning demonstrates the utility of phase space methods in harmonic
analysis of automorphism groups \cite{Arveson}. One may therefore expect that they will be of 
further use in this interesting field.

\bigskip

\noindent{\bf Acknowledgements:}
This work is a part of a joint project with Prof. D. Buchholz to whom I am grateful for many valuable suggestions,
especially for communicating to me the proof of Lemma~\ref{time}. Financial support from Deutsche Forschungsgemeinschaft 
is gratefully acknowledged.

\begin{appendix}
\section{Proof of Theorem \ref{Nregions} }
\setcounter{equation}{0} 
The argument is based on the following  abstract lemma:
\begin{lemma}
\label{key} Let $X$ and $Y$ be Banach spaces, $S_k\in X^*$ for $k\in\{1,\ldots,N\}$ and
$\tau\in Y$ be s.t. $\|\tau\|=1$. Then the  $\eps$-content of the map
$\The: X\to Y\ot\compsup$ given by
\beq
\The(\fun)=\tau\, (S_1(\fun),\ldots, S_N(\fun)),\quad \fun\in X, \label{standard}
\eeq
satisfies the bound
\beq
\N(\eps)_{\The}\leq (4eN)^{\fr{2^7\pi \|\The\|_2^2}{\eps^2}}, \label{standard1}
\eeq
where $\|\The\|_2=\sup_{\fun\in X_1}(\sum_{k=1}^N|S_k(\fun)|^2)^{\fr{1}{2}}$.
\end{lemma}
\proof Fix $\eps>0$ and let $\lattice_0=\{ (n_1+in_2)\eps \ | \ n_1,n_2\in\integer\}$. For each
$k\in\{1,\ldots, N\}$ and $\fun\in X_1$ we choose $J_k(\fun)\in \lattice_0$ so that
$|S_k(\fun)-J_k(\fun)|\leq \sqrt{2}\eps$ and $|J_k(\fun)|\leq |S_k(\fun)|$. Define
the set $\lattice=\{J_1(\fun),\ldots,J_N(\fun) \ | \ \fun\in X_1 \}$ of all $N$-tuples 
appearing in this way. We claim that $\#\lattice\geq \N(4\eps)_{\The}$. In fact, assume that there  
are $\fun_1,\ldots,\fun_K\in X_1$, $K>\#\lattice$, s.t. for $i\neq j$ there holds
\beq
4\eps<\|\The(\fun_i)-\The(\fun_j)\|
=\sup_{k\in\{1,\ldots, N\}} |S_k(\fun_i)-S_k(\fun_j)|.
\eeq
Then there exists such $\khat$, depending on $(i,j)$, that $4\eps<|S_{\khat}(\fun_i)-S_{\khat}(\fun_j)|$.
Consequently, by a $3\eps$-argument
\beq
|J_{\khat}(\fun_i)-J_{\khat}(\fun_j)|
\geq|S_{\khat}(\fun_i)-S_{\khat}(\fun_j)|-2\sqrt{2}\eps>\eps,
\eeq
which shows that there are at least $K$ different elements of $\lattice$ in contradiction
to our assumption.

In order to estimate the cardinality of the set $\lattice$ we define
$M=\big[\fr{\|\The\|_2^2}{\eps^2}\big]$, assume for the moment that $0<M\leq 2N$ and denote by 
$V_M(R)\leq e^{2\pi R^2}$ the volume of the $M$-dimensional ball of radius $R$. Then
\beqa
\#\lattice\leq\sum_{\su{n_1,\ldots,n_{2N}\in\integer \\ n_1^2+\cdots+n_{2N}^2\leq M} } 1 \leq
\binom{2N}{M} 2^M V_{M}(2\sqrt{M})\leq (4Ne)^{8\pi M}. \label{ball}
\eeqa
We note that each admissible combination of integers $n_1,\ldots, n_{2N}$ contains at most $M$ 
non-zero entries. Thus to estimate the above sum we pick $M$ out of $2N$ indices and consider the points 
$(n_{i_1},\ldots, n_{i_M})\in\integer^M$ which  belong to the  $M$-dimensional ball of radius $\sqrt{M}$.
Each such point is a vertex of a unit cube which fits into a ball of radius $2\sqrt{M}$ (since $\sqrt{M}$
is the length of the diagonal of the cube). As in $M$ dimensions a cube has $2^M$ vertices, there
can be no more than $2^M V_{M}(2\sqrt{M})$ points $(n_{i_1},\ldots,n_{i_M})\in\integer^M$ satisfying the 
restriction $n_{i_1}^2+\cdots+n_{i_M}^2\leq M$. In the case $M\geq 2N$ a more stringent bound
(uniform in $N$) can be established by a similar reasoning. For $M=0$ there obviously holds $\#\lattice=1$.
\qed\\
\emph{ Proof of Theorem \ref{Nregions}. } Fix $0<p<\fr{2}{3}$. Then Condition $\Nnat$ provides, for any $\de>0$, 
a decomposition of the map $\PiE$ into rank-one mappings $\Pi_n(\,\cdot\,)=\tau_n\,S_n(\,\cdot\,)$, where
$\tau_n\in\mfa(\mco)^*$ and $S_n\in\trac^*$,
s.t. 
\beq
(\sum_{n=1}^{\infty}\|\Pi_n\|^p_{\x})^{\fr{1}{p}}\leq (1+\de)\|\PiE\|_{p,\x}.
\eeq
Assuming that the norms $\|\Pi_n\|_{\x}$ are given in  descending order with $n$, we obtain the bound
\beq
\|\Pi_n\|_{\x}\leq \fr{(1+\de)\|\PiE\|_{p,\x}}{n^{1/p}}. \label{xxnorm}
\eeq 
Similarly, we can decompose the map $\Theh$ into a sum of maps $\The_n$ of the form
\beqa
\The_n(\fun)=\big(\Pi_n(\al_{x_1}^*\fun),\ldots,\Pi_n(\al_{x_N}^*\fun) \big)
            =\tau_n\big(S_n(\al_{x_1}^*\fun),\ldots,S_n(\al_{x_N}^*\fun)).
\eeqa
Now we can apply Lemma \ref{key} with $\tau=\tau_n/\|\tau_n\|$ and $S_k(\,\cdot\,)=\|\tau_n\|S_n(\al_{x_k}^*\cdot\,)$.
From estimate (\ref{xxnorm}) we obtain
\beqa
\|\The_n\|_2=\sup_{\fun\in\tracB}(\sum_{k=1}^N\|\tau_n\|^2|S_n(\al_{x_k}^*\fun)|^2)^{\fr{1}{2}}&=&\|\Pi_n\|_{\x}\non\\
&\leq& \fr{(1+\de)\|\PiE\|_{p,\x}}{n^{1/p}}.
\eeqa
Substituting this inequality to the bound (\ref{standard1}) we get 
\beq
\N(\eps)_n\leq (4eN)^{\fr{2^7\pi(1+\de)^2\|\SiEh\|_{p,\x}^2}{\eps^2 n^{2/p}}}. \label{factor}
\eeq
We conclude with the help of Lemmas 2.3 and 2.4 from \cite{BA} that the $\eps$-content of the map $\Theh$ satisfies
\beq
\N(\eps)_{E,\vxb}\leq \prod_{n=1}^\infty \N(\eps_n)_n \label{product}
\eeq
for any sequence $\{\eps_n\}_1^\infty$ s.t. $\sum_{n=1}^\infty\eps_n\leq \fr{\eps}{4}$. We choose
$\eps_n=\fr{\eps}{4} \fr{n^{-\fr{2}{3p}}}{\sum_{n_1=1}^\infty n_1^{-\fr{2}{3p}}}$,
make use of the bounds (\ref{factor}) and (\ref{product}), and take the infinum w.r.t. $\de>0$. There follows
\beq
\N(\eps)_{E,\vxb}\leq (4eN)^{\fr{2^{11}\pi\|\SiEh\|_{p,\x}^2 
 }{\eps^2}(\sum_{n=1}^\infty n^{-\fr{2}{3p}} )^3 }.
\eeq
With the help of Condition $\Nnat$ we obtain the bound in the statement of Theorem~\ref{Nregions}. \qed

\section{Proof of Theorem \ref{sharp}}
\setcounter{equation}{0}
Since the expansion of $\PiE$ into rank-one maps which appears in  Theorem \ref{sharp}
differs slightly from those which are considered in the existing literature
\cite{BP,Bos3}, we outline here the construction.\\
\emph{ Proof of Theorem \ref{sharp}. }
First, we recall from \cite{Bos3} Sect. 7.2.B. that given any pair of 
multiindices $\mub=(\mup,\mum)$
and an orthonormal sequence of $J$-invariant vectors (e.g. $\{e_j\}_1^{\infty}$), there
exist weakly continuous linear functionals $\phi_{\mub}$ on $\mfa(\mco)$ s.t.
\beq
\phi_{\mub}(W(f))=e^{-\half\|f\|^2}\lan e|\fp\ran^{\mup}\lan e|\fm\ran^{\mum},
\eeq 
which satisfy the bound
\beq
\|\phi_{\mub}\|\leq 4^{|\mub|}(\mub!)^\half, \label{stateestimate}
\eeq
where $\mub!=\mup!\mum!$. These functionals can be constructed making use of the equality
\beqa
(\vac|[a(e_1),[\ldots,[a(e_k),[a^*(e_{k+1}),[\ldots, [a^*(e_l),W(f)],\ldots]\vac)\non\\
=e^{-\half\|f\|^2}\prod_{n_1=1}^k\lan e_{n_1}| if\ran \prod_{n_2=k+1}^l\lan if| e_{n_2}\ran.
\eeqa
Next, we evaluate the Weyl operator on some $\fun\in\trac$, rewrite it in a normal ordered form
and expand it into a power series
\beqa
&\fun&\!\!\!\!\!(W(f))\non\\
&=&e^{-\fr{1}{2}\|f\|^2}
\sum_{m^{\pm},n^{\pm}\in\nat_{0}}\fr{i^{m^++n^++2m^-}}{m^+!m^-!n^+!n^-!}
\fun(a^*(\fp)^{m^+}a^*(\fm)^{m^-}a(\fp)^{n^+}a(f^-)^{n^-}).\qquad
\eeqa
Subsequently, we expand each function $f^\pm$ in the orthonormal basis $\{e_j\}_1^{\infty}$ of $J$ invariant eigenvectors of the operator $\Sa$:
$f^\pm=\sum_{j=1}^{\infty}e_j\lan e_j|f^\pm\ran$. Then, making use of the multinomial formula, we obtain
\beq
a^*(\fp)^{m^+}=\sum_{\mu^+,|\mu^+|=m^+}\fr{m^+!}{\mu^+!}\lan e|\fp\ran^{\mup} a^*(\Lp e)^{\mu^+}
\eeq
and similarly in the remaining cases. Altogether we get
\beqa
\fun( W(f) )=\sum_{\mub,\nub} \fr{i^{|\mu^+|+|\nu^+|+2|\mu^-|} }{\mub!\nub!}\phi_{\mub+\nub}(W(f))
\fun(a^*(\LL e)^{\mub}a(\LL e)^{\nub}) \non\\
=\sum_{\mub,\nub}\tau_{\mub,\nub}(W(f))\T_{\mub,\nub}(\fun),\label{weylexp}
\eeqa
where $\tau_{\mub,\nub}(\,\cdot\,)=\fr{i^{|\mu^+|+|\nu^+|+2|\mu^-|}}{\mub!\nub!}\phi_{\mub+\nub}(\,\cdot\,)$. 
We recall that in the massive case
$\T_{\mub,\nub}=0$ if $|\nub|>\M$ or $|\mub|>\M$. Consequently, for the relevant indices there holds
\beq
\|\tau_{\mub,\nub}\|\leq \fr{4^{|\mub|+|\nub|} }{(\mub!\nub!)^{\h} }\bigg(\fr{(\mub+\nub)!}{\mub!\nub!}\bigg)^{\h}\leq 2^{\fr{5}{2}(|\mub|+|\nub|)}\leq 2^{5\M},
\eeq 
where we made use of the bound (\ref{stateestimate}) and properties of the  binomial coefficients. Now it 
follows from estimate (\ref{traces}) that for any $p>0$
\beqa
& &\sum_{\mub,\nub}\|\tau_{\mub,\nub}\|^p\|\Tmn\|^p\leq 2^{5p\M}\M^{p\M} (\sum_{k=0}^{[\M]}\|\Sa^p\|_1^k)^4. 
\eeqa
In view of this fact and of weak continuity of the functionals $\tau_{\mub,\nub}$, equality (\ref{weylexp}) can be extended to any $A\in\mfa(\mco)$. In other words
\beq
\PiE(\fun)(A)=\fun(A)=\sum_{\mub,\nub}\tau_{\mub,\nub}(A) \T_{\mub,\nub}(\fun),
\eeq
what concludes the proof of the theorem. \qed
\end{appendix}

\end{document}